# Perspectives on Privacy in the Post-Roe Era: A Mixed-Methods of Machine Learning and Qualitative Analyses of Tweets

Yawen Guo, MISM[1], Rachael Zehrung, MS[1], Katie Genuario, BS[1], Xuan Lu, PhD[2],
Qiaozhu Mei, PhD[2], Yunan Chen, PhD[1], Kai Zheng, PhD[1]
[1] University of California, Irvine, Irvine, CA, USA; [2] University of Michigan, Ann Arbor, MI, USA

**Abstract**

*Abortion is a controversial topic that has long been debated in the US. With the recent Supreme Court decision to overturn Roe v. Wade, access to safe and legal reproductive care is once again in the national spotlight. A key issue central to this debate is patient privacy, as in the post-HITECH Act era it has become easier for medical records to be electronically accessed and shared. This study analyzed a large Twitter dataset from May to December 2022 to examine the public's reactions to Roe v. Wade's overruling and its implications for privacy. Using a mixed-methods approach consisting of computational and qualitative content analysis, we found a wide range of concerns voiced from the confidentiality of patient–physician information exchange to medical records being shared without patient consent. These findings may inform policy making and healthcare industry practices concerning medical privacy related to reproductive rights and women's health.*

**Introduction**

The issue of abortion and the legal framework surrounding it has been a controversial topic in the US for many years. During the colonial era, regulations varied across different colonies based on religious and legal traditions.[1] Societal attitudes towards abortion shifted in the early 1900s, resulting in the stigmatization of medical practitioners who performed abortions, primarily women.[2] In 1973, the landmark Supreme Court decision in Roe v. Wade recognized abortion as a constitutional right, stating that women seeking an abortion have a right to privacy.[3] The decision was upheld in 1992 in Planned Parenthood of Southeastern Pennsylvania v. Casey. However, on June 24, 2022, the Supreme Court overturned Roe v. Wade, which paved the way for individual states to curtail or outright ban abortion rights, bringing the topic back in the national spotlight.

Privacy is a fundamental right protected by the US Constitution. In the context of the abortion debate, it refers to the right to control one's personal information and decisions related to reproductive health without undue interference or surveillance from the government or other entities.[5] Scholars have also identified additional dimensions of privacy that are specific to reproductive health. For example, Warren and Brandeis defined privacy as the "right to be let alone"[5] and Cohen argued that the privacy interests of women seeking an abortion include not only bodily autonomy and decision making, but also the right to access safe and confidential medical care.[6] Privacy anxieties therefore lie at the heart of the abortion discourse and could considerably shape public opinion.[7]

The potential impact of overturning Roe v. Wade is significant and far-reaching, encompassing not only privacy rights, but also specific concerns related to technology and medical privacy. The widespread adoption of electronic health records in US hospitals and clinics as a result of the 2009 HITECH Act has created more opportunities for data breaches and made it easier than before to access sensitive medical information, including by the government and law enforcement. The decades-long effort in building health information exchange infrastructures to promote data sharing across healthcare organizations further exacerbates this concern. Patients often have little knowledge or control over who has access to their medical records, leading toward doubts about privacy and confidentiality, particularly on women's reproductive health decisions.[8]

While traditional survey methods have been used to gauge public opinion on a range of issues, they are cost-prohibitive to conduct and may not provide an accurate reflection of the public's views, especially on sensitive matters. Therefore, social media platforms such as Twitter have become a valuable source of information for policy makers and healthcare and public health practitioners to understand the public's sentiments on controversial issues. Social media platforms provide a convenient venue for people to express their views, opinions, and concerns that may not be voiced otherwise. They also allow for real-time collection of large volumes of data at relatively low cost. To date, several studies have analyzed social media discussions related to the overruling of Roe v. Wade. Chang et al. were the first to curate a substantial dataset of Twitter posts pertaining to the abortion rights debate and presented preliminary results on topics such as opinion dynamics/ polarization and protest mobilization.[9] Two recent studies by Fan et al. and Mane et al. conducted sentiment analyses of Twitter data from May 1 to mid-July 2022 and revealed that over half of the tweets conveyed a neutral sentiment while the remaining exhibited a predominantly negative tone.[10, 11] However, to date, very few studies have specifically investigated social media discussions on privacy concerning the overturn of Roe v. Wade. To address the gap, we used a mix method of computational and qualitative content analysis to analyze a large Twitter dataset collected from May to December 2022. This timeframe covers several key milestones during the re-energized abortion debate from the Supreme Court draft decision being leaked to the public on May 2nd to the aftermath of the 2022 midterm election in November. The objective of this study was to answer the following three research questions (RQs):

> **RQ1**: What are the main topics being discussed on Twitter regarding the overrunning of Roe v. Wade?
>
> **RQ2**: How has the volume and topics of the Twitter discussions on Roe v. Wade evolved before and after the official Supreme Court overruling?
>
> **RQ3**: What are the specific viewpoints, interpretations, concerns, and attitudes expressed by Twitter users regarding privacy in relation to the overruling?

**Methods**

**Data collection**

The empirical data were collected over an eight-month period from May 2 to December 31, 2021. This time period can be divided into three distinct phases each marked by a major milestone, namely the leak of the Supreme Court's draft decision on May 2nd, official overruling of Roe vs. Wade on June 24th, and the 2022 US midterm election on November 8th. Table 1 shows more detail on each of these phases.

**Table 1.** The three distinct phases of the study.

| Phase | Dates | Milestone |
| --- | --- | --- |
| Decision Leak | 5/2–6/23/2022 | The leak of a draft of the Supreme Court majority decision in Dobbs v. Jackson Women's Health Organization that would explicitly overturn Roe v. Wade. There was widespread speculation during this period that led to furious protesting campaigns. |
| Official Overruling | 6/24–11/7/2022 | Release of the official overruling of Roe v. Wade that effectively nullified the landmark 1973 decision. The legal and political battles that followed during this period were centered on the enforcement and implementation of the new ruling. |
| Electoral Impact | 11/8–12/31/2022 | The 2022 midterm election had a significant impact on the future of abortion rights in the US. The election results were closely watched by both sides of the debate and had the potential to shape the direction of policy and legal efforts. |

The data were retrieved using the Twitter Decahose API that provides a 10% random sample of all tweets circulated.[12] For each tweet, we obtained the timestamp indicating the time at which the tweet was posted, an identifier, and the text of the tweet itself. For retweets, we included the text of the original tweet. For tweets quoting another tweet, we included both the text of the quoted tweet and the added comment in the tweet.

We used an iterative process to obtain a comprehensive corpus of relevant discussions on Twitter. We started with amassing a substantial corpus of tweets covering a wide breadth of content potentially relevant to the topic. Subsequently, we undertook a detailed analysis of the retrieved tweets to further refine the keyword list in order to ensure a balance on inclusiveness and relevance. This analysis arrived at 3 main topics of "Roe V. Wade", "prolife", and "prochoice", which accordingly led to 10 pairs of keywords: {roe}*{wade}, and {abortion}*{roe, prolife, pro-life, pro life, right to life, righttolife, prochoice, pro-choice, pro choice}. Specifically, a selected tweet should contain at least one pair of the keywords to be included in the analytical dataset. For example, a tweet with the snippet "The Supreme Court will likely overturn Roe v. Wade" will be selected because it contains both "Roe" and "Wade".

**Text classification to remove content containing no expression of personal opinions**

A substantial proportion of the tweets collected, although containing keywords related to the topic, lacked an expression of personal opinions concerning the overturn of Roe v. Wade. For instance, many tweets were merely for sharing news articles reporting on the decision leak or the overruling of Roe v. Wade. As this study aims to examine public opinion and discourse on reproductive rights and privacy, it is essential to eliminate irrelevant content from the curated data. To achieve this, we developed a supervised machine-learning model to identify tweets that are both pertinent to the topic and contain personal opinions. The inclusion of retweets in classification could potentially compromise accuracy, as it may bias the analysis towards the context of the original tweet rather than that of the user.[13] Moreover, incorporating retweets could introduce extraneous data and obscure the genuine opinion expressed in the user's own tweets. Thus, we excluded retweets from further analyses.

To prepare the training data, three authors (KG, RZ, and YG) manually annotated a random sample of tweets. For each tweet, the authors evaluated whether it was relevant to the overruling of Roe v. Wade and, if so, whether the tweet contained personal opinions on the target topic. The authors first coded 100 tweets independently to calibrate the results. The Cohen kappa interrater reliability[14] was 0.70 between YG and KG, 0.69 between YG and RZ, and 0.55 between KG and RZ. After consensus development meetings, the authors independently coded another set of 100 tweets, achieving an interrater reliability of 0.92 between YG and KG, 0.90 between YG and RZ, and 0.89 between KG and RZ. The three authors then proceeded to annotate 900 additional tweets, dividing the workload equally among them. The final training dataset used to train the supervised model thus contained a total of 1,100 tweets.

In addition, a set of preprocessing steps were applied to all tweets including converting all text to lowercase, removing punctuations and hashtags, and performing stemming.[15] Afterwards, a 100-dimensional vector representation was generated for each word using the word2vec algorithm.[15] In this study, we comparatively assessed the performance of three machine-learning models including logistic regression,[16] support vector machine,[17] and Long Short-Term Memory (LSTM) Network,[18] which are commonly employed in similar text classification tasks. These machine-learning models were programmed using tensorflow libraries[19] in Jupyter Notebook 4.8.[20]

**Computational topic modeling on longitudinal evolution of Twitter discussions**

To examine the longitudinal evolution of public discourse surrounding the overruling of Roe v. Wade, we conducted a topic modeling analysis of weekly tweets across the three distinct phases (i.e., draft decision leak, official overruling, and electoral impact). For the data in each phase, we studied the top features using Tf-idf [21] and top words using Latent Dirichlet Allocation,[22] a popular topic modeling approach, to identify commonly discussed topics and how they evolved over time during the eight-month study period.

**Qualitative content analysis**

Following the topic modeling, we conducted a qualitative content analysis using the grounded theory approach.[23] The purpose was to gain a more in-depth understanding of Twitter users' attitudes, interpretations, viewpoints, and concerns regarding the prospect of patient privacy in the aftermath of Roe v. Wade's overruling. We performed a dictionary-based search on the overall Twitter corpus to narrow down to specific tweets that contained relevant discussions on privacy.[24]

To develop the dictionary, we studied existing literature, including scoping reviews, systematic reviews, and research articles, to identify patient-related privacy topics [25–27] and consumers' discussions about privacy on social media.[28,29] Additionally, we explored the mention of privacy in the literature specifically concerning the legal framework surrounding abortion and the relevant court cases.[30–32] The resulting dictionary is reported in Table 2, which includes a general category "privacy" to ensure maximum coverage of all relevant content. Based on the dictionary, we searched for relevant tweets, requiring each identified tweet to contain at least one keyword.

Then, we sampled a random set of 100 tweets for open coding to generate the initial qualitative schema. Three coders (KG, RZ, and YG) independently coded the data. Discrepancies were resolved through consensus development meetings. Each tweet was coded with a maximum of three codes. While theoretical saturation was achieved after coding approximately 70 tweets, we went on to code an additional set of 900 tweets (300 for each coder) in order to enable follow-up quantitative analysis (e.g., to calculate the proportion of tweets that supported a particular viewpoint).

**Table 2.** Dictionary for identifying tweets with privacy relevance.

| Category | Dictionary |
| --- | --- |
| privacy | "privacy", "private", "consent", "confidential" |
| data privacy | "cybersecurity", "encryption", "hipaa", "track", "app", "digital", "personal data", "personal information", {"data", "information"}*{"security", "breach", "protect"} |
| medical record | "EHR", "EMR", {"health", "medical", "hospital", "patient", "clinical"}*{"record", "information", "document", "data"} |
| data sharing | {"data", "information", "record"}*{"share", "sharing", "exchange"}, {"cross"} * {"border", "state"}, "disclose", "disclosure", "out of state", "outside state" |

**Results**

Using the initial set of search keywords, we retrieved a total of 867,546 tweets during the eight-month study period. After removing retweets, 178,562 unique original tweets remained. LSTM, with a dropout layer of 0.1 to prevent overfitting, exhibited the best performance among the models considered for the first step of removing irrelevant content.[33] We trained the LSTM model on a dataset of 990 tweets and evaluated it on a separate dataset of 110 tweets. Its overall accuracy on the validation set was 83.84% and the F-1 score was 90.23%. Out of the 178,562 original tweets, 118,550 were deemed relevant and were used in subsequent analyses.

Figure 1 shows the weekly volume of relevant tweets. The first two spikes correspond to around May 2, 2022, when the Supreme Court's majority decision was leaked; and around June 24, 2022, when Roe v. Wade was officially overturned. In November leading toward the midterm election, the volume of relevant tweets picked up again but at a much milder magnitude.

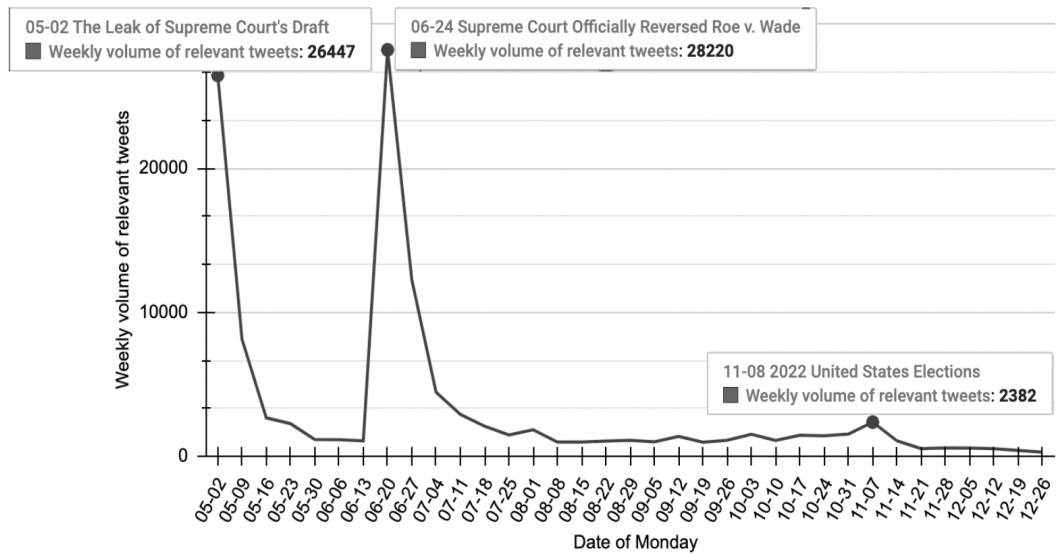

**Figure 1.** Change of tweet volume over time

The results of topic modeling are reported in Table 3. Upon dividing the corpus into three distinct phases and testing with different cluster sizes, we determined that a 3-cluster model produced the most meaningful topic separation. Twitter discussions during the first phase (decision leak) focused on legality of abortion at the state versus the federal level. In the second phase (official overruling), the topics shifted toward the political landscape surrounding reproductive rights. In the third phase (electoral impact), the focus shifted again back to legality of abortion, in addition to morality. Across all three phases, the debate remained highly contentious and polarized, with various vocal actors seeking to shape the public discourse around federal versus state law, the Supreme Court's role in policy making, and the rights of the unborn fetus versus women's bodily autonomy.

**Table 3.** Evolution of frequently discussed topics

| Phase | Representative Words of a Topic |
|---|---|
| Decision leak | Women's rights, constitutional ban, state laws, privacy issue, abortion |
| | Pro-choice vs pro-life, rape, pregnancy, murder, children, control, support, care |
| | Supreme Court decision, Democrats vs Republicans, overturning Roe v. Wade, state versus federal law |
| Official overruling | Women's rights, constitutional ban, state laws, privacy issue, abortion |
| | Biden's stance, Congress, government power, codifying Roe v. Wade, Supreme Court justices |
| | Medical control, pregnancy, rape, children, support, care, pro-abortion vs anti-abortion |
| Electoral impact | Abortion as murder, women's rights, care, support, pro-abortion versus anti-abortion |
| | Election and political maneuvering, social issues, codifying Roe v. Wade |
| | Trump's influence, federal vs state law, Supreme Court decision, women's rights, overturning |

Using the dictionary-based approach, we identified a total of 3,107 tweets that might contain relevant discussions on privacy. A significant portion of these tweets however only mentioned the general right to privacy without providing any substantial detail in the context of Roe vs. Wade overruling. Therefore, we further eliminated tweets that only contained the keywords "right" and "privacy", which reduced the size of the corpus to 1,674. We then performed qualitative coding on a random sample of 1,000 tweets, of which 354 were further deemed to be irrelevant. Among them, some words from the dictionary appeared only in the user's name, or certain words were included in the tweet but were not indicative of a discussion related to privacy. Table 4 summarizes the qualitative themes that emerged from the data and their frequency of occurring across the 646 tweets analyzed.

**Table 4.** Themes from the qualitative content analysis.

| Theme | Sub-theme | Description | Count | Sample tweet |
|---|---|---|---|---|
| Concern | Political and legal distrust | Distrust of political or legal institutions and their decision-making processes | 59 | *"I keep waiting for the Dems to splint the privacy law that lifted Roe. It wouldn't even be an abortion bill, it would be a patient privacy law. But apparently it's not worth it."* |
| | Prosecution and criminalization | Concern that individuals may face prosecution or criminalization for seeking or providing abortions | 75 | *"With Roe V. Wade being overturned, it's important folks' advocacy is abolitionist in nature. Make no mistake, people will be jailed for self managed abortion, crossing state lines to receive abortion care, which has already started."* |
| | Invasion of privacy | Concern that individuals' privacy may be violated | 37 | *"When Roe v. Wade falls, Google's location stockpile will be weaponized against abortion seekers. Tell Google to stop collecting and storing this data to protect reproductive rights."* |
| | Wealth, power, and equity | Concern that access to reproductive care may be limited by wealth or power differentials | 19 | *"Poor women will be more affected by the Roe V Wade decision. Others can afford to cross state lines. So there will be more women with babies on welfare."* |
| | Impact on women and women's rights | Concern for the impact of restricting abortion on women's rights and bodily autonomy | 73 | *"I denounce the Supreme Court's decision to overturn Roe v. Wade. It marks a sad chapter in American history. The health of thousands of women across many states will be jeopardized with this ruling."* |
| | Shared with the government or law enforcement | Concern that personal information about abortion may be made available to the government or law enforcement | 27 | *"What is unclear at this time is whether any information on those individuals would be shared with American law enforcement or state officials looking to prosecute people for traveling for an abortion, as some experts have suggested could be next if Roe v. Wade falls."* |
| | Shared with employer or insurance company | Concern that personal information about abortion may be made available to employers or insurance companies | 26 | *"I hope all that are smiling got that same smirk on your face when you end up having to a pay full price for a medical tragedy and you lose your job when your boss gets your records and doesn't want to accommodate THIS is Roe v wade"* |
| Privacy Right | Core of Roe v. Wade | Acknowledgement that the core of Roe v. Wade is the right to privacy | 180 | *"You know Roe vs Wade was about privacy, not abortions?"* |
| | Right to privacy | Impact on general privacy rights including gun, marriage, contraception, transgender, and social security. | 162 | *"Right to privacy was a huge part of Roe V Wade. And honestly, conservatives have been railing against the privacy basis for abortion since Roe was decided. To use it now for guns is a huge middle finger to liberals."* |
| | Constitutionality | Privacy rights' constitutional validity and codification | 246 | *"Killing an innocent child has nothing to do with privacy. RGB even stated Roe v Wade would be overturned because it really did not pass the Constitutional test."* |
| Medical Privacy | General medical privacy | The broader concept of privacy related to medical procedures | 88 | *"Roe v Wade is about medical privacy. Abortion was the subject. The door is open for government/state input into all medical decisions, for everyone."* |
| | Medical record data shared | Concern for medical records being shared without consent | 9 | *"Your medical records are now public records that can be requested by anyone, without your consent. That was a big part of Roe v Wade."* |
| | Physician disclosure/patient doctor confidentiality | The confidentiality of information shared between patients and physicians | 14 | *"What happened to doctor patient confidentiality? Putting this info online means it can be hacked. They might use it to see if anyone's had an abortion & use the info to charge ppl with crimes."* |
| | Bodily autonomy and medical decision making | Individuals' rights to bodily autonomy and medical decision-making | 85 | *"Abortion is a matter of privacy and bodily autonomy. There is no negotiating the terms. We have places here in the US where it's legal to pray for your child instead of seeking medical help. We are not pro-life, we are pro-suffering."* |
| Information Privacy | General information privacy | Privacy related to information and data in the digital age | 62 | *"Roe v Wade is way bigger than abortion. It's more about protected health information than anything. Republicans claim they want less government overreach but the Supreme Court just removed a barrier to protected health information and privacy."* |
| | Personal data tracking and sharing | Personal data being tracked and shared without consent | 57 | *"I go to sleep and wake up to find out that my period tracker is a government spy that can be used against me in court because of roe v wade being overturned."* |
| | Privacy recommendation | Suggestions on how to protect individuals' privacy | 25 | *"Please make the switch from Flo to Clue. It's a European app, it's safe to use and our data won't be given away. With the overturning of Roe vs Wade we need to be diligent in protecting ourselves"* |

*All sample tweets included in the table and the rest of this paper were paraphrased to protect the identity of the user

**General privacy rights**

Through the qualitative content analysis, we discovered that a substantial proportion of users upheld the notion of the right to privacy in general. However, their viewpoints diverged concerning the significance of overturning Roe v. Wade in a privacy context, the legal validity of the ruling, and the ramifications of the decision on other civil liberties. Many users perceived the essence of Roe v. Wade to be privacy-related, implying that revoking the ruling could potentially result in the loss of other rights with similar constitutional foundations, such as marriage. One user conveyed this sentiment by stating that "*Their long game was to overturn Roe v Wade, which leads to no more privacy laws, no contraception, no gay marriages, and no interracial marriages. They've been saying this stuff for decades.*" Another perspective commonly expressed was that while privacy rights remain crucial, abortion constitutes an exceptional case concerning codification. As one user articulated, "*Privacy is a different issue, yes, it's been the basis of Roe for 50 years, but what they're attempting is to codify abortion as a Right unto itself. I also believe we should codify privacy issues, but they aren't the same.*"

**Medical privacy**

In terms of medical privacy, users questioned how the overturn of Roe v. Wade would impact the confidentiality of the patient–provider relationship and medical decision making, as well as broader implications for bodily autonomy in a medical context. Given the potential legal ramifications for patients seeking abortions as well as healthcare providers who perform abortion services, users expressed concern that physicians would be compelled to disclose protected health information (PHI) to law enforcement. Some users believed that this disclosure could result in legal prosecution, even for miscarriages or medically necessary abortions.

Concerns around the dissolution of patient–provider confidentiality extended beyond the possibility of prosecution, as users considered how different actors might misuse their PHI. As one user explained, "*I'm concerned about Right to Privacy between a patient and doctor going down in Roe V Wade - as someone who takes a very expensive medication ($12,000/month), what happens if that information is available to employers or insurers?*" (T62). With access to their medical records, users worried that insurance companies and employers would be able to discriminate based on medical conditions (e.g., "*you lose your job when your boss gets your records and doesn't want to accommodate,*" T81). They questioned whether the removal of Roe v. Wade could open the door to more widespread governmental interference in personal health matters, such as vaccination mandates.

Many users highlighted parallels between decision making and bodily autonomy for reproductive health, gender-affirming care, and vaccinations. With the removal of protections provided by Roe v. Wade, some users believed that "*an individual could be forced to get a vaccine even if they don't want to*" (T149). On the other hand, we observed tweets suggesting that bodily autonomy never existed, and therefore an overturn would not have a significant impact on medical decision making. For instance, V99 questioned, "*what's the big deal (Roe v Wade) ... folks across the country lost their right to informed consent/ autonomy with respect to deciding whether they want the Covid 19 vaccine.*" These viewpoints were typically associated with users who expressed political distrust in the Democratic party, and believed that pro-choice proponents were hypocritical in their support for medical privacy. As T97 stated in a reply, "*People like you didn't support private healthcare decisions when it came to masks and shots, and forced people to do both as a condition of employment, education and purchase.*" Abortion and vaccination mandates are both politically polarizing topics, which users compared to contextualize the discussions around bodily autonomy.

**Information privacy**

In terms of data and information privacy in a digital era, users expressed concerns about the potential for their personal data to be accessed and misused by various entities. For example, with recent political developments such as the Supreme Court's decision to overturn Roe v. Wade, users worried that law enforcement officials may seek to collect personal data for prosecutions in states that have criminalized abortions. As one tweet noted, "*With the Supreme Court's decision that legalized abortion, collected location data, text messages, search histories, emails and period and ovulation-tracking apps could be used to prosecute people who seek an abortion — or medical care.*"

Users also expressed concerns about the state of data protection and privacy laws in the US, particularly as related to personal data tracking apps. They worried that such apps could be used to criminalize personal decisions or gather evidence for prosecution (e.g., "*does anyone have any period tracking methods that won't potentially get me in trouble due to Roe v Wade? I've never had an abortion or been pregnant but I do have an inconsistent period and sometimes will miss mine.*"). In response, some users recommended using alternative methods to store and track personal health information such as apps developed in European countries, which they believed to be more protective of user data due to European Union's more stringent data protection laws: "*If you're worried, use Clue because it's German and can't sell data.*" The issue of information privacy also extended to the potential for data sharing involving other types of technology companies. For example, one tweet stated, "*When Roe v. Wade falls, Google's location stockpile will be weaponized against abortion seekers. Tell Google to stop collecting and storing this data to protect.*"

**Discussion**

The objective of this study was to examine public attitudes expressed on Twitter regarding privacy in the context of overruling the post-Roe era. Using a mixed methods approach consisting of machine learning and qualitative analysis, we identified public opinions and the underlying rationales on general privacy rights, medical privacy, and information privacy. We found that Twitter users expressed concerns over the loss of privacy and continentality protection following the overturn of Roe v. Wade, particularly related to healthcare decisions and personal data sharing. Based on our findings, we emphasize the significance of safeguarding medical and information privacy in the post-Roe era, and discuss implications for patients, healthcare providers, and policy makers.

Patients' concerns around privacy and the strength of the patient–physician relationship can significantly impact their information-sharing behavior with healthcare providers,[34] especially when it comes to sensitive information such as reproductive health.[35] Similar to prior work,[36] we found that users worried about the sharing of specific types of information and the potential misuse of their PHI by different parties such as healthcare organizations, insurance companies, and the government. Users also questioned the legal implications of Roe v. Wade for patient–provider confidentiality protection. Patients might be hesitant to disclose PHI due to concerns around providers sharing information with law enforcement without patient consent, which could result in prosecution.

While this study focuses on the perspectives of the general public, we also draw attention to the implications of the overruling for healthcare providers and their relationships with patients. Providers should prioritize building trust and establishing open communication with their patients to encourage health information exchange in order to make informed patient care decisions. In the post-Roe era, it is particularly important for providers to address patient concerns about privacy and confidentiality; ensuring that patients are aware of their rights and protection under HIPAA; and communicating the circumstances under which they would be required to disclose patients' PHI to other entities. Even though HIPAA protects PHI to some extent, providers are still required to release patient information to comply with court orders or subpoenas, which could be issued to investigate abortions. Providers and healthcare facilities thus have to navigate a complex legal landscape and protect themselves (e.g., if they provide abortion services), while ensuring the privacy of patients' sensitive information. The potential legal ramifications could also impact providers' documentation practice, as they might want to minimize the amount of information recorded when documenting reproductive health encounters.[37] This new paradigm of documentation practice, while contributes to improved privacy and confidentiality protection, could lead to potential issues of omitting important medical history information that undermines quality of care. Future work therefore should consider engaging directly with clinicians who provide reproductive care services to investigate their perceptions around privacy and confidentiality, as well as strategies they may adopt to protect patients' rights. Further, many patients may be unaware of the specific abortion laws in their state, highlighting the importance of healthcare providers' role in helping their patients make informed decisions.

In addition to concerns around patient–provider confidentiality, our study also shows Twitter users highlighted their hesitance to use personal health applications such as period tracking apps, which have become increasingly popular to enhance their ability to record and use personalized health information. However, with sensitive information such as reproductive health data, it is important for these apps to prioritize privacy protection for their users. One possible recommendation for these apps is to make users aware of the reproductive health laws in their state and disclose their role in complying with investigative requests. Additionally, these apps could implement strong data privacy and security measures, such as end-to-end encryption and secure storage of user data. Providing clear and transparent privacy policies and options for users to control their data sharing could also help to build trust and encourage continued use of these apps. By prioritizing privacy protection for their users, personal health applications can help to ensure the safe and secure use of sensitive health data in the post-Roe era. Legislators may also consider enhancing the regulation to provide oversight on how such apps use and share user data.

This study has several limitations. First, Twitter users who are willing to share their personal opinions on this sensitive issue publicly are by no means representative of the broader population. Therefore, the findings of this study, and of other social media-based studies more generally, need to be interpreted with caution. Second, our data retrieval methods may not be inclusive enough to capture all relevant discussions. The keywords-based search strategy could omit infrequently used words and phrases expressing opinions on the topic of interest, in addition to opinions expressed in non-textual forms (e.g., images, videos). Third, with the advancement of technologies based on regenerative language models, it has become increasingly difficult to differentiate Twitter content posted by social bots versus by human users. Research using social media data, particularly tweets, could thus be susceptible to manipulation by special interest groups that attempt to dictate the narrative in a public discourse. Future research thus needs to incorporate more robust bots detection mechanisms especially when studying conversational topics that may have significant policy implications and societal impact.

**Conclusion**

Privacy is a crucial topic when it comes to the discussion of reproductive rights in the US. This study may shed light on the complex and nuanced perspectives of the public regarding privacy in the context of the abortion debate. The use of a mixed methods approach allowed us to obtain a comprehensive understanding of privacy concerns ranging from broad privacy rights to medical and information privacy. The results highlight the need for more effective public health communication and the restoration of trust between the public and the government. Addressing these concerns is vital for ensuring access to reproductive care and facilitating productive dialogue on this controversial issue.